**Concave/convex switchable lens using active phase-change material Ge$_3$Sb$_2$Te$_6$**

Xueliang Shi, Juan Liu, Gaolei Xue, Weiting Peng, Bin Hu, Yongtian Wang

*Beijing Engineering Research Center for Mixed Reality and Advanced Display, School of Optics and Photonics, Beijing Institute of Technology, Beijing, 100081, China*

**Abstract**
Normally, the focal length of a conventional lens is fixed. Scientists have made much effort in modulating it into bifocal, which is very important for virtual reality (VR) and argument reality (AR) 3D display. It is even much more difficult for a lens to realize both convex and concave functions with one geometric structure, that is, a concave/convex switchable lens, which can tune 3D real-images and virtual-images in AR and VR, corresponding to long depth of view in 3D display. Based on the tunable refractive indexes of phase-change materials, here we propose a series of concave/convex switchable lenses. When the phase-change material is in different states, one switchable lens is able to perform a negative or positive focal length, or perform negative and positive focal lengths simultaneously. The lenses can be either cylindrical, spherical or other types. For the superior characteristics, these switchable lenses can be employed in various optical fields.

In geometrical optics, the modulation of a lens is attributed to the refraction of light at the lens surfaces. However, the theory of wave optics suggests that it is caused by the phase modulation. The refractive index of the glass or other transparent lens materials is larger than that of the surrounding air, which leads to the slowing down of wave front. For a convex lens, the slowing effect of light waves passing through the center of the lens is greater than that of the light waves passing through both sides of the lens, resulting in the incident light wave convergence. On the contrary, a concave lens diverges the incident light. However, the focal length of a conventional lens is fixed. Usually we need to adjust the optical systems by mechanical modulation or other methods [1].

Scientists have made much effort in realizing a lens with bifocus, which is very important in intraocular lens[2], 3D display[3], target tracking[4], and other fields[5]. It is even much more difficult to realize a lens that can switch between convex lens functions and concave lens functions with one geometric structure. This lens can be widely applied in tuning of 3D real-images and virtual-images in AR and VR, corresponding to long depth of view in 3D display.

This switchable device may be realized by phase-change materials, such as In$_2$Se$_3$ [7], Bi$_2$Te$_3$[8], VO$_2$[9], and GeSbTe[10] which can switch between crystalline and amorphous states, and the corresponding physical properties are different.[6] In this paper, we propose a series of concave/convex switchable lenses based on phase-change materials. When the phase-change material is in one state, the lens is a concave lens; and in the other state, it is a convex lens.

assuming the substrate material is air ($n_0$=1). Each wave zone is composed of two rectangular stages constructed from phase-change material with different heights. Other structures of wave zones are similar, with the same height and different widths. The final result is shown in **Figure 1**a. Switched to concave/convex lens, all the rectangular structures in Figure 1a should be in the same state.

We then conduct simulations to verify the conversion effect of the structure in Figure 1a in the x-z plane. The simulations are carried out using finite-difference time-domain solutions by Lumerical Solutions. Starting from the center of the lens, there are 16 wave zones on both sides. The heights of the two stages in each wave zone are $\lambda/2$, $3\lambda/2$ respectively. The phase-change material is assumed to be $Ge_3Sb_2Te_6$. The simulated incident light is parallel light ($\lambda=3.1$ μm). The results are shown in figure 1b and c.

When the phase-change material is in amorphous state ($n_1=3.5+0.001i$), the structure exhibits the function of a convex lens with a focal length f '=50 μm, as shown in Figure 1b. When the material is in crystalline state ($n_2=6.5+0.06i$), as shown in Figure 1b, the lens reflect the light to one point. Due to the large refractive index, most of the incident light energy is reflected or absorbed within the lens. The reflected light is convergent, indicating that the lens is a concave lens with a focal length f '= -50 μm. It is also seen from Figure 1b and c that the focusing of the light beam is not very satisfactory. One reason is that, we design the structure based on a spherical lens, which have dispersion and other aberrations. On the other hand, each wave zone has only two rectangular stages. If more stages in each wave zone are used, the accuracy of the lens can be improved.

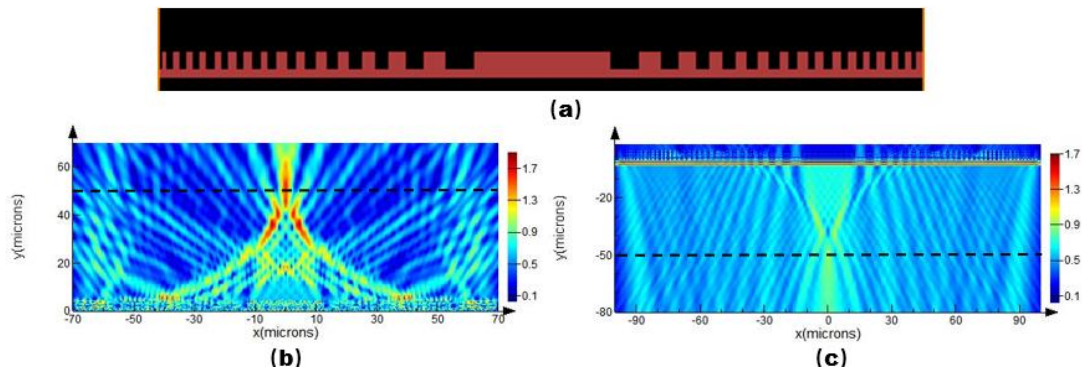

**Figure 1**. a) A switchable lens, with a two-stage structure without substrate material. The lens simulated is composed of the phase-change material $Ge_3Sb_2Te_6$). b) The result when the lens is in amorphous state, exhibiting the function of a convex lens. c) The result when the lens is in crystalline state, exhibiting the function of a concave lens.

In conclusion, utilizing phase-change materials with excellent optical properties to change the focal length of the lens, a series of concave/convex switchable lenses are proposed. To sum up, the switchable lenses we proposed have simple structure and can applied to large optical components, which are suitable for practical processing. Our next job is to put forward more reasonable lenses, and to make an actual lens for practical experiments. However, in order to produce a high-precision switchable lens, we need to further find suitable materials as the substrate mentioned in this article.

**Funding**
National Natural Science Founding of China (NSFC) (61575024, 61420106014), and the UK Government's Newton Fund.